\documentclass[12pt]{article}

\usepackage{newtxtext,newtxmath}

\usepackage{graphicx}

\usepackage[letterpaper,margin=1in]{geometry}

\renewenvironment{abstract}
	{\quotation}
	{\endquotation}

\date{}

\renewcommand\refname{References and Notes}

\makeatletter
\renewcommand{\fnum@figure}{\textbf{Figure \thefigure}}
\renewcommand{\fnum@table}{\textbf{Table \thetable}}
\makeatother

\usepackage{scicite}

\usepackage{url}

\def\scititle{
	Attosecond pulses from a solid driven by a synthesized two-color field at megahertz repetition rate
}
\title{\bfseries \boldmath \scititle}

\author{
	Zhaopin~Chen$^{1,2,3\ast\dagger}$,
	Mark~Levit$^{1,2,3\dagger}$,
	Yuval~Kern$^{1,2,3}$,
    Basabendra~Roy$^{1,2,3}$\and
	Adi~Goldner$^{1,2,3}$,
	Michael~Kr\"uger$^{1,2,3\ast}$\and
	\small$^{1}$Department of Physics, Technion---Israel Institute of Technology, Haifa 32000, Israel.\and
	\small$^{2}$Solid State Institute, Technion---Israel Institute of Technology, Haifa 32000, Israel.\and
 	\small$^{3}$The Helen Diller Quantum Center, Technion---Israel Institute of Technology, Haifa 32000, Israel.\and
	\small$^\ast$Corresponding authors. Email: zhaopin.chen@campus.technion.ac.il, krueger@technion.ac.il\and
	\small$^\dagger$These authors contributed equally to this work.
}

\begin{document} 

\maketitle

\begin{abstract} \bfseries \boldmath
Probing coherent quantum dynamics in light-matter interactions at the microscopic level requires high-repetition-rate isolated attosecond pulses (IAPs) in pump-probe experiments. To date, the generation of IAPs has been mainly limited to the kilohertz regime. In this work, we experimentally achieve attosecond control of extreme-ultraviolet (XUV) high harmonics in the wide-bandgap dielectric MgO, driven by a synthesized field of two femtosecond pulses at 800\,nm and 2000\,nm with relative phase stability. The resulting quasi-continuous harmonic plateau with $\sim 9\,\mathrm{eV}$ spectral width centered around 16.5\,eV photon energy can be tuned by the two-color phase and supports the generation of an IAP ($\sim 730$ attoseconds), confirmed by numerical simulation based on three-band semiconductor Bloch equations. Leveraging the high-repetition-rate driver laser and the moderate intensity requirements of solid-state high-harmonic generation, we achieve IAP production at an unprecedented megahertz repetition rate, paving the way for all-solid compact XUV sources for IAP generation. 

\end{abstract}

\noindent
\section*{Introduction}

High-harmonic generation (HHG) produces coherent extreme ultraviolet (XUV) light through the interaction of strong laser pulses with gases~\cite{Ferray1988,McPherson1987}, solids~\cite{Ghimire2011,Vampa2015a} and liquids~\cite{Luu2018} on a table-top platform. In the time domain, isolated attosecond pulses (IAPs) can be achieved, which are key to the extremely precise pump-probe measurements of attosecond science~\cite{Corkum2007,Krausz2009,Calegari2016}. Starting with the first successful observation of an IAP in a gas in 2001~\cite{Hentschel2001}, various techniques for producing IAPs from atomic gases have been established~\cite{Chini2014}, mostly based on spectral filtering in conjunction with an approach to effectively reduce the number of HHG-generating half-cycles to a single one~\cite{Christov1997}, such as polarization gating~\cite{Sansone2006}, double optical gating~\cite{Hiroki_PRL2010}, ionization gating~\cite{Ferrari_NP2010} and two-color mixing~\cite{eiji_NC2013}. Another frontier in HHG is the generation of attosecond pulses at high repetition rates of 100\,kHz and higher, which is beneficial for experiments where only low single-shot signal rates can be achieved, such as photoemission spectroscopy and imaging~\cite{Heinrich2021}, or for precision spectroscopy with frequency combs in the XUV~\cite{Porat2018a}. High repetition rates are essential for improving measurement efficiency, minimizing space charge effects, preventing sample damage, and enhancing overall signal quality. An early study produced IAPs in a gas at a repetition rate of 600\,kHz~\cite{Manuel_NP2013}, which is, to the best of our knowledge, the highest repetition rate for IAPs.

In principle, HHG from solids~\cite{Ghimire2011,Ghimire2019,Goulielmakis2022} opens up the possibility to generate IAPs more efficiently than in gaseous media due to the higher atomic density in solids~\cite{Ndabashimiye2016} and the non-parabolic dispersion of the band structure, which enables strong enhancement in certain spectral regions~\cite{Uzan2020a}. The first demonstration of an IAP from a solid has been achieved with synthesized sub-cycle light field transients generated from a hollow-core fiber supercontinuum at kHz repetition rates~\cite{Garg2016,Garg2018}. Bulk pre-compression in a solid enables such transients, opening the door to attosecond pulses~\cite{Hammond2017}. Solid-state HHG sources in the XUV allow for a more modest peak intensity of the driving laser pulses, on the order of $10^{13}\,\mathrm{W\,cm}^{-2}$ or less~\cite{Ghimire2014,Garg2016, Hammond2017}, due to the fact that the band gaps of materials like ZnO, MgO and SiO$_2$ are smaller than the usual ionization energies of noble gases. These properties make solid-state HHG particularly attractive for high repetition rates, alleviating the requirements on the driving laser system. Solid-state HHG has been achieved at tens of MHz oscillator repetition rates using a titanium-sapphire laser in a metal-sapphire nanostructure~\cite{Han2016} and using a mid-infrared (MIR) laser in bulk ZnO~\cite{Vampa2019}. A first pump-probe spectroscopy application of attosecond pulses generated in ZnO has recently been reported at 100\,kHz repetition rate~\cite{Nayak2024}. IAP generation from solids, however, has remained elusive at MHz repetition rates.

In this work, we employ a laser field synthesized from two incommensurate frequencies and demonstrate high-contrast IAP generation from a wide bandgap dielectric at a repetition rate of 1\,MHz. Compared to other incommensurate HHG experiments~\cite{Bruner2018,Bruner2021}, the two constituent fields at 800\,nm and 2000\,nm wavelength are phase-stable with respect to each other and form the synthesized laser field with controllable waveform on the attosecond scale. We employ a magnesium oxide (MgO) crystal for HHG due to its wide bandgap, high cutoff energy and spectral enhancements~\cite{Ghimire2014, You2017, Uzan2020a, Uzan2022}. These properties facilitate the generation of IAPs. Moreover, MgO exhibits a comparatively high damage threshold, thus allowing us to access the extreme ultraviolet (XUV) spectral range. Through attosecond waveform control, we observe an HHG spectrum with greatly reduced oscillatory features. Our measurement is in excellent agreement with numerical simulations. While we do not directly measure the temporal profile of the generated XUV light field, our calculations show that a high-contrast IAP with a duration of 730\,as at a center photon energy of 16.5\,eV is produced.

\section*{Results}

\subsection*{Synthesized two-color field for high-harmonic generation}

\begin{figure}[!htb]
\centering
\includegraphics[width=0.7\textwidth]{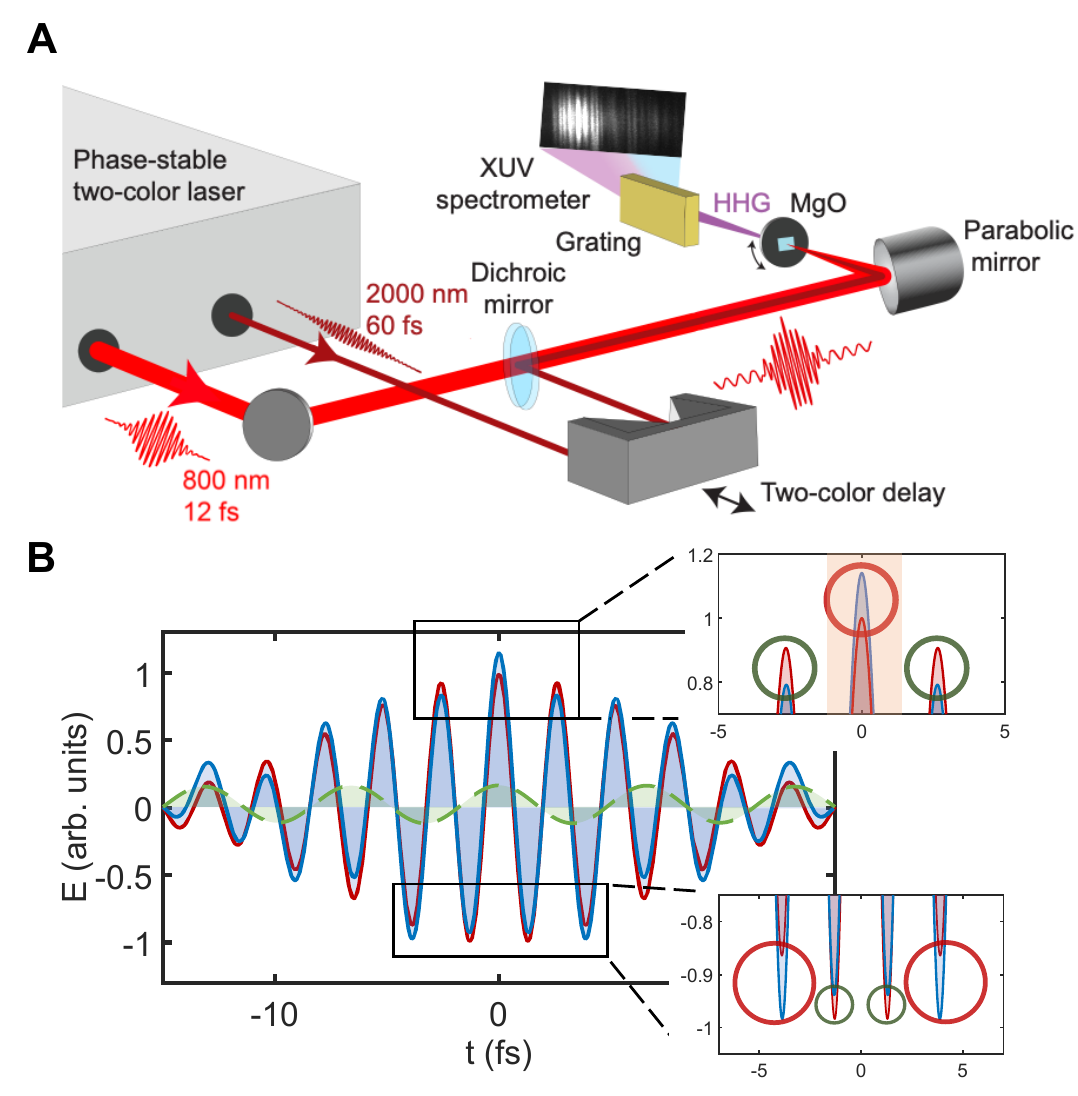}
\caption{{\bf Experimental setup and synthesized two-color field distribution.} ({\bf A}) Experimental setup for attosecond control of high-harmonic generation and isolated attosecond pulse generation from MgO with synthesized two-color light fields. ({\bf B}) The field profile (solid blue with shading) shows a dominant field peak in the center. Red solid (with shading) and green dashed curves (with shading) depict the 800\,nm and the 2000\,nm field, respectively. The peak intensity of 2000\,nm field is $2\%$ of the 800\,nm peak intensity. The red and green circles in the insets indicate where the 2000\,nm field enhances and suppresses the corresponding half-cycle, respectively.}\label{fig1}
\end{figure}

In our experiment (see Fig.~\ref{fig1}{A} for a sketch of the setup), we generate high harmonics and implement attosecond control by synthesizing two-color laser pulses using a phase-stable optical parametric chirped pulse amplification (OPCPA) system operating at a repetition rate of 1\,MHz (Class 5 Photonics White Dwarf~\cite{Braatz2021}). The OPCPA system outputs two distinct pulses, 12\,fs pulses at a wavelength of 800\,nm and 60\,fs pulses at 2000\,nm. Mathematically, the synthesized two-color field can be expressed as $E(t)=A_{1}\exp(-2\ln{2}\,t^2/\tau_1^2)\cos{\left(\omega_{1}t+\phi_\mathrm{CEP,1}\right)}+A_{2}\exp(-2\ln{2}\,(t+\Delta t)^2/\tau_2^2)\cos{\left(\omega_{2}t +\phi +\phi_\mathrm{CEP,2} \right)}$, where the subscripts 1 and 2 denote the 800\,nm and 2000\,nm laser pulses, and $\Delta t$ and $\phi$ are the two-color delay time and phase ($\phi=\omega_2\Delta t$). The carrier-envelope phases (CEPs) $\phi_\mathrm{CEP,1}$ and $\phi_\mathrm{CEP,2}$ are passively stabilized at an unknown fixed value. We assume the CEPs to be zero for both colors since they barely affect our results (see Supplementary Text and Figures S4 and S5 for a discussion of the CEP effects). We emphasize the fact that in our case, the two wavelengths are incommensurate, meaning they cannot be related by an integer factor. In our experiment, we maintain the peak intensity of the 800\,nm pump at approximately $\mathrm{15\,TW\,cm^{-2}}$ at the focus. We then introduce a weak admixture of the second color, at a relative intensity of 2\%. The relative delay between the two colors is controlled by a translation stage with attosecond precision. The resulting synthesized pulse waveform at zero two-color delay is depicted in Fig.~\ref{fig1}{B}. A key characteristic of the synthesized field is the presence of a strongly enhanced field peak in the orange shaded area at the pulse center, while the adjacent field peaks of the 800\,nm pulse at $t\approx 1.33 \mathrm{fs}$ and $t\approx 2.66 \mathrm{fs}$ are suppressed, indicated by the green circles, due the constructive or destructive interference of the two fields. The intensity ratio of the synthesized field between the central peak and the strongest adjacent side peak, at around $\pm4\,\mathrm{fs}$, is about $72\%$. Even in the case with a CEP for both pump pulses equal to $\pi/2$, the intensity ratio $79\%$ is still sufficient for a good contrast (see Fig.~S3). The synthesized pulse is akin to a single-cycle pulse (see Fig.~\ref{fig1}A), which enables us to produce a quasi-continuous plateau in the HHG spectrum. Due to the extremely nonlinear nature of the process in play, this ratio can already give rise to a dominant attosecond pulse with large intensity contrast to adjacent pulses, thus equivalent to an IAP. 

To optimize the HHG yield, we align the polarization of the laser field along the $\Gamma-X$ orientation of the MgO crystal. This orientation gives rise to a $\mathrm{7.5\,eV}$ bandgap at the $\Gamma$ point and a $\mathrm{\sim 17\,eV}$ bandgap at the $X$ point between the valence band and first conduction band. A strong spectral enhancement is located at the $X$ point bandgap~\cite{Uzan2020a}, which we leverage in our experiment. In addition, with sufficiently intense field strength, the second conduction band may also be involved and contribute to higher energy harmonics, up to $\sim 25\,$eV~\cite{TaoYuan2018}. With this wide bandgap, an XUV spectrum with sufficient spectral width supporting an IAP is feasible. We generate XUV high harmonics in the transmission geometry within the last $\sim 10\,$nm of the crystal and measure the spectrum using a flat-field XUV spectrometer based on a grating and a microchannel plate (MCP) detector (see Fig.~\ref{fig1}A for a sketch and Materials and Methods for further details).

\subsection*{Attosecond control of solid-state high harmonics}

\begin{figure}[!htb]
\centering
\includegraphics[width=0.9\textwidth]{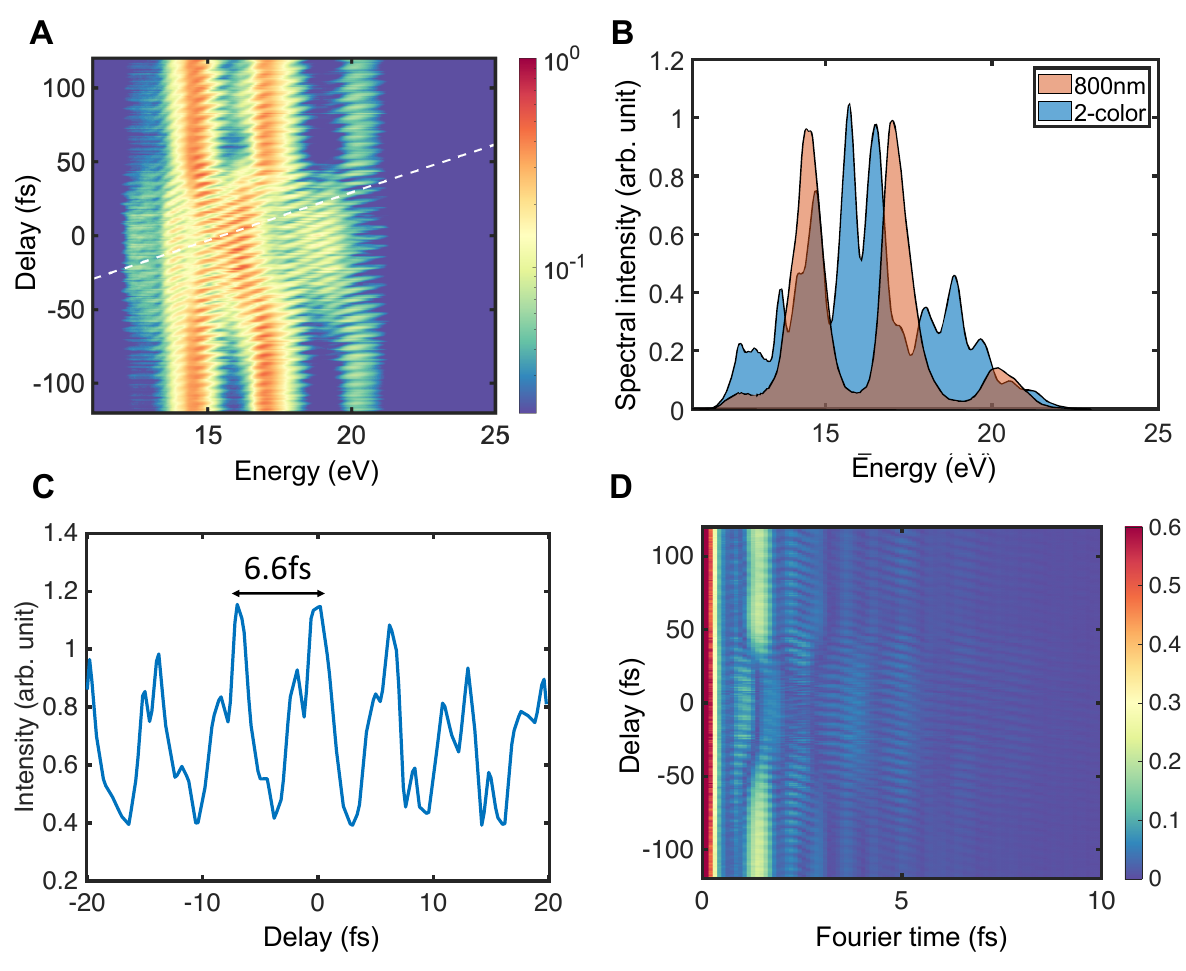}
\caption{ {\bf Attosecond-controlled harmonic spectra}. ({\bf A}) Experimentally measured harmonic intensity as a function of two-color delay from $-120\, \mathrm{fs}$ to $120\, \mathrm{fs}$. The white dashed line has a slope of $6.5\,\mathrm{fs}\,\mathrm{eV}^{-1}$. ({\bf B}) Experimentally measured harmonic intensity for the 800\,nm pulse pump (orange) and for the synthesized two-color field (blue) with $0.5\, \mathrm{fs}$ delay. ({\bf C}) Spectral intensity at $15.5\,\mathrm{eV}$ as a function of two-color delay shows a period of $6.6\,\mathrm{fs}$. ({\bf D}) Harmonic intensity in the Fourier time domain as a function of delay, corresponding to ({\bf A}).
}\label{fig2}
\end{figure}

We begin by comparing the high-harmonic (HH) spectra generated by the 800\,nm field alone with those induced by the synthesized two-color field, as shown in Fig.~\ref{fig2}A and B. The HH spectra driven solely by the 800\,nm field exhibit odd harmonic peaks with a spacing of $2\hbar \omega_{0}=3.1\,\mathrm{eV}$, where $\omega_{0}$ is the fundamental frequency. Due to the approximately 4.5 cycles contained within the 12\,fs duration of the 800\,nm pulse, the resulting HHG spectrum shows discrete odd harmonics. At a peak intensity of $\sim 15\,\mathrm{TW\,cm^{-2}}$, the measured HH spectra exhibit a cutoff energy around $21\,\mathrm{eV}$. Since the MCP detector in our XUV spectrometer is only efficient for photon energies larger than $\sim 11\,$eV (see Fig.~S6), our experimental measurement shows this part of the spectrum. 

Introducing the weak 2000\,nm laser pulse disrupts the symmetry of the original field, resulting in the occurrence of a dominant peak field at a specific delay. It thus leads to a nearly continuous spectrum with spectral width of $\sim 9$\,eV around the spectral singularity at $\sim 17$\,eV. Figure~\ref{fig2}A shows the measured HH spectra as a function of two-color delay, which is the central result of our work. When the relative delay of two pulses are larger than half of the pulse duration of the 2000\,nm, discrete harmonic spectra are mainly produced by the 800\,nm pulse. In contrast, the temporal overlapping of the two pulses within a time delay from $-30\,\mathrm{fs}$ to $30\,\mathrm{fs}$ leads to a quasi-continuous spectrum, with several lower contrast peaks with a spacing of $\sim 1$\,eV (see also Fig.~\ref{fig2}B). This already indicates the spectral interference originating from a dominant attosecond pulse and weak subordinate pulses at a temporal spacing of $\sim 4\,\text{fs}$.

Scanning the delay between the 800\,nm and 2000\,nm fields reveals a periodic modulation in the HHG spectrum, as shown in Fig.~\ref{fig2}A. The spectral interference forms a tilted stripe-like pattern with a slope of approximately 6.5\,fs eV$^{-1}$. A cross-section of the delay-dependent spectral intensity at 15.5\,eV in Fig.~\ref{fig2}C highlights a periodic modulation with a period of 6.6\,fs. This modulation arises from the temporal distribution of the synthesized two-color field. For incommensurate continuous light waves, the synthesized field can be expressed as
$E(t, \phi) = A_{1} \cos{(\omega_{1} t)} + A_{2} \cos{(\omega_{2} t + \phi)}$, where the field naturally repeats after delays corresponding to the optical periods of each of the two colors. However, for an ultrashort pump pulse consisting of only a few cycles, this repetition behavior no longer holds. The 12\,fs duration of the 800\,nm pulse is too short to approximate it as a continuous wave. On the other hand, the 60\,fs 2000\,nm pulse can be reasonably treated as a plane wave within the temporal overlap region. As a result, the constructive overlap of the two fields' peaks occurs periodically, repeating only after one optical cycle of the 2000\,nm pulse (see Fig.~\ref{fig3}E).

To better understand the change in harmonic spectra from the non-overlap to the overlap region of our two-color field, we perform an inverse Fourier transformation of the spectral intensity into the Fourier time domain, assuming a uniformly zero spectral phase. Although the Fourier time domain does not give us access to the actual spectral phase and the exact temporal structure of each attosecond pulses produced in each laser shot, it allows us to extract other crucial information about the train structure of the attosecond pulses. All attosecond pulses contributing to the HHG spectrum and their timing cause interference in the spectral domain and the Fourier transformation can reveal temporal spacings between the attosecond pulses as they are modulated using the two-color delay, in analogy to a HHG spectral interferometry scheme utilizing the CEP~\cite{Ott2013}. Out of temporal overlap, the Fourier amplitude plot of the harmonic spectra shows a secondary peak at a Fourier time of approximately $\tau \sim 1.3$\,fs. This peak corresponds to temporal features separated by half a cycle of the 800\,nm pulse, showing that a train of attosecond pulses is generated and rendering IAP generation elusive. Remarkably, this changes completely in the overlap region. The secondary peak at $1.3$\,fs is suppressed due to the enhancement of the electric field in the dominant half-cycle and the suppression of the electric field in the adjacent half cycles (see Fig.~\ref{fig1}B for a comparison of the electric fields of the 800\,nm pulses and the two-color pulses). Additionally, in the overlap region, we find an enhancement of the peak around $\tau \sim 4$\,fs, which corresponds to a temporal delay of attosecond pulses of three optical half-cycles of the 800\,nm field. This is the result of constructive interference between the dominant attosecond pulse generated around the peak of the field and one or more weak satellite pulses at a temporal distance of 4\,fs.

\subsection*{Comparison with numerical simulations}

To elucidate our experimental findings, we utilize the semiconductor Bloch equations (SBEs) to predict and calculate solid-state HHG (see Materials and Methods for more details). Generally, both interband polarization and intraband current acceleration contribute to HHG in solids. In our case, the HHG yield in MgO is mainly contributed by the interband polarization according to our numerical calculations (see Fig.~S7), in accordance with prior works~\cite{Uzan2020a,Uzan2022}. The analogy between the interband polarization process in solids and the semi-classical three-step model in atomic HHG indicates the possibility of generating IAPs in MgO.

\begin{figure}[htb!]
\centering
\includegraphics[width=0.92\textwidth]{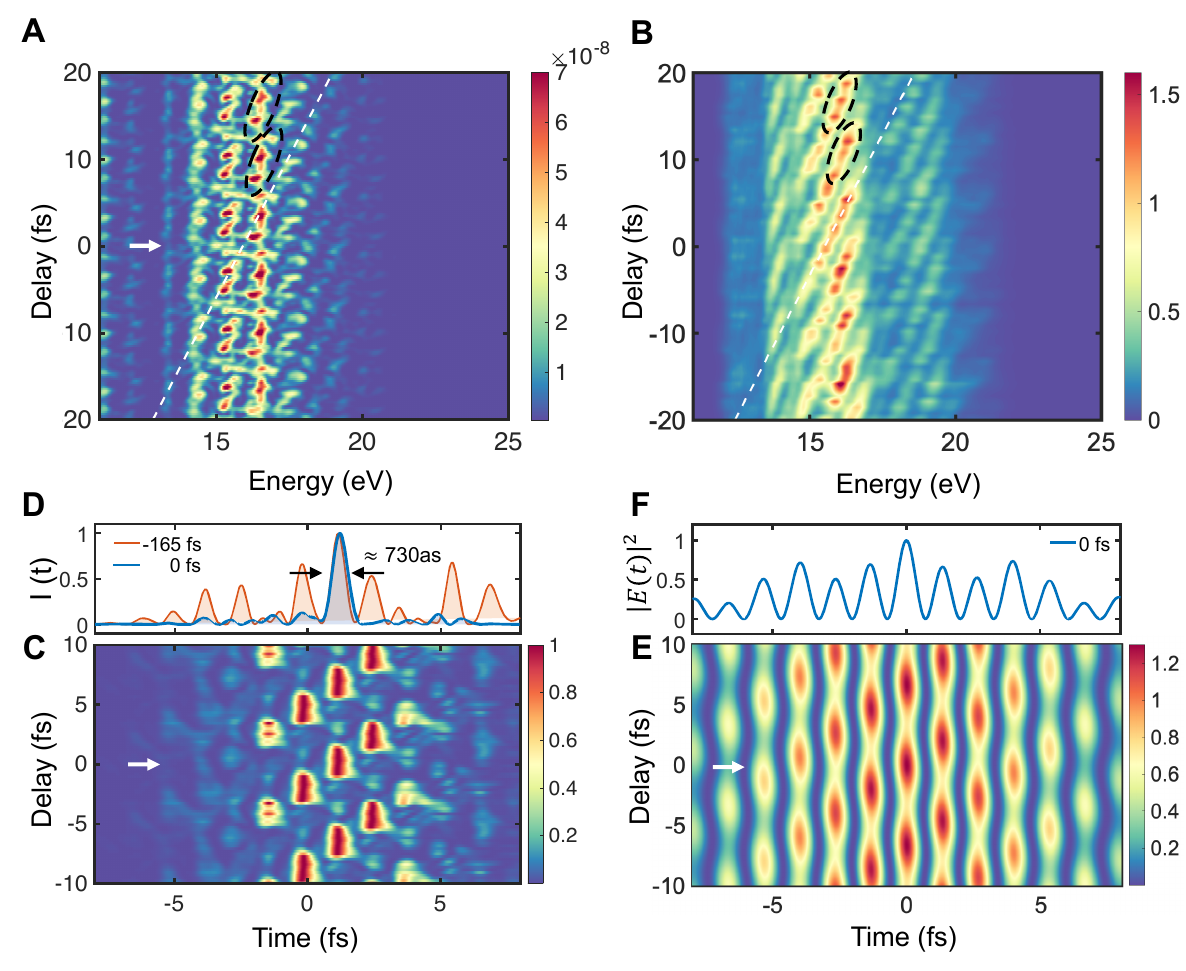}
\caption{{\bf Comparison of experimental results with SBEs simulation results.} Semiconductor Bloch equations calculation and experiment results. ({\bf A}) and ({\bf B}) Harmonic intensity as a function of two-color delay for SBE simulation and experiment (logarithmic scale), respectively. For both cases, the peak intensity for the 800\,nm pulse is $\mathrm{15\,TW\,cm^{-2}}$.
The black dashed circle curves shows three local maxima for each 6.6\,fs period for both experiment and simulation. The white dashed line for both cases have a slope of $6.5\, \mathrm{fs}\,\mathrm{eV^{-1}}$.  The white arrow in ({\bf A}) indicates zero delay. ({\bf C}) Attosecond pulses intensity as a function of two-color delay in real time. The white arrow indicates zero delay where a high-contrast IAP is produced. ({\bf D}) Comparison of the formation of attosecond pulse trains and IAPs at time delay $-165\,\mathrm{fs}$ (orange) and $0\,\mathrm{fs}$ (blue), respectively. ({\bf E}) Field strength square distribution $|E(t)|^{2}$ of the synthesized pulse as a function of the two-color delay. The white arrow indicates zero delay. ({\bf F}) shows the synthetic field strength square distribution with a two-color delay equal to $0\, \mathrm{fs}$. }\label{fig3}
\end{figure}

The numerical simulations of the SBEs in Fig.~\ref{fig3}A perfectly match the experimental results in Fig.~\ref{fig3}B. Here, the two-color delay is taken from $-20\, \mathrm{fs}$ to $20\, \mathrm{fs}$ in the overlap region between two pump pulses. Similar to the experimental results, our numerical simulation also displays a 6.6\,fs periodicity in the spectral interference as a function of the time delay. A tilted stripe-like spectral interference pattern shows a slope of $\sim 6.5\,\mathrm{fs\,eV^{-1}}$, exactly corresponding to the experimental one. 
Since the harmonics peak shift linearly from one order to the next within one cycle of the 2000\,nm field, the slope can be expressed as $T_{2\mathrm{um}}/\Delta E = 6.5\,\mathrm{fs\,eV^{-1}}$, where $T_{2\mathrm{um}}$ is the duration of an optical cycle of the 2000\,nm field and $\Delta E$ is the period of the spectral interference from the generated attosecond pulses. Therefore, we can obtain the mean spectral period as $
\Delta E \sim 1.0 \, \text{eV}$, which already indicates the interference is mainly from two attosecond pulses with a temporal distance of $\sim$ 4\,fs, corresponding to 1.5 cycles of the 800\,nm field.  According to established theory of gas-phase and solid-state HHG~\cite{Pascal1998,Ott2013,ShaLi2023}, the phase of attosecond pulses, governed by the harmonic dipole phase, depends on the field strength and the time spent by the electron and hole in the conduction and valence bands, respectively. Since the two-color delay periodically changes the field strength in each half-cycle (see the squared absolute value of the synthesized field distribution in Fig.~\ref{fig3}E), the relative harmonic phases contributed by attosecond pulses in different half-cycles vary correspondingly. As a consequence, the spectral interference also shifts as the relative harmonic phases from different attosecond pulses change. 

Taking a closer look at the delay dependence of the spectra, a surprising result is that there are fine oscillatory structures within each $6.6\,\mathrm{fs}$ period, showing three local maxima of the harmonic intensity within that period for both simulation and experiment (see Fig.~\ref{fig3}A and B). These local maxima originate from the spectral interference of attosecond pulses, indicating the existence of more than one significant attosecond pulse at a specific two-color delay. To verify this assumption, we further use the numerical results to estimate the real-time distribution of the attosecond pulses in Fig.~\ref{fig3}C. Here, we can see that the attosecond pulses in real time show a periodic distribution and repeat itself for each 6.6\,fs of the two-color delay. Within each period, there are three dominant pulses at different two-color delays, along with weak subordinate pulses. The local maxima observed in the spectrum originate from the interference between two dominant attosecond pulses located at the "switch" positions. Since the generation of attosecond pulses is strongly related to the field strength within each half cycle, we expect that the synthesized two-color field also has a similar temporal distribution. Indeed, the synthesized field strength square distribution $|E(t)|^2$ in Fig.~\ref{fig3}E contains three local maximum values located at at $\left( \phi, t\right)  =\left(0, 0\right)$, $\left(\pi- \pi\omega_2/\omega_1, \pi/\omega_1 \right)$ and $\left(\pi+ \pi\omega_2/\omega_1, -\pi/\omega_1\right)$ for the two-color delay within one cycle of the second color. These electric field local maxima give rise to comparatively more continuous spectra located between the local spectral maxima. Compared with the synthesized field distribution, the half-cycle delay of the attosecond pulse generation indicates that electrons and holes in MgO generated at the original half-cycle will re-collide with each other and emit harmonics at the next half cycle. 
Further comparison between experiment and simulation regarding the large range of two-color delay in both spectral and Fourier time domain can be found in the Supplementary Information.

\section*{Discussion}

The good match between the numerical and experimental results further validates our theoretical model. Therefore, we can use the numerical results to estimate the real-time distribution of the attosecond pulses. For the experimental realization of an IAP, a filter that acts as a spectral window to remove the fundamental light and lower-order harmonics is required. Here, we consider an indium filter~\cite{henneck1997}, which has a transparency window between 12 and 21\,eV photon energy. To compare the attosecond pulse distribution generated by one-color and two-color fields, we examine the intensity distribution of the generated attosecond pulse at two specific delay positions (see Fig.~\ref{fig3}D). When the delay time is equal to 0\,fs, an IAP with a full width at half maximum (FWHM) duration of $730\,\mathrm{as}$ is generated. At a delay of $-165$\,fs, where two pump pulses are widely separated, an attosecond pulse train is generated due to the discrete high-harmonic spectra. By controlling the two-color delay, we can switch between an IAP and a pulse train and use their corresponding spectral signatures as a guide.

In conclusion, we demonstrate attosecond control of HHG from the dielectric MgO, driven by a combination of two femtosecond pulses with incommensurate wavelengths and controllable relative phase. By precisely controlling the temporal overlap of the two-color field, we synthesize the equivalent of a single-cycle pulse, producing a quasi-continuous XUV spectrum without the need for complex single-cycle femtosecond pulse compression techniques. This result is confirmed by our theoretical model based on the semiconductor Bloch equations, showing the creation of an IAP. By combining the advantages of a two-color optical parametric chirped pulse amplification system -- featuring high repetition rates, phase stability, and ultrashort pulse durations -- alongside the moderate intensity requirements of solid-state HHG, we achieve IAP generation at 1\,MHz repetition rate. With photon energies in the lower XUV range ($\sim$10 eV to $\sim$20 eV), our high-repetition-rate IAPs are ideally suited for pump-probe experiments, enabling the study of attosecond electron dynamics in wide bandgap dielectrics such as LiF~\cite{LiF_PNAS2008}, deep valence-to-conduction band excitations in solid materials~\cite{Martin_Science2014}, and molecular charge migration~\cite{Calegari_Science2014}.

\section*{Materials and methods}

\subsection*{Semiconductor Bloch equations calculations}

To describe the interaction of the synthesized two-color field and the MgO crystal in the $\Gamma$-$X$ orientation, we use the following one-dimensional, three-band semiconductor Bloch equations (SBEs):

\begin{equation}\label{SBE3_P}
\begin{aligned}
i \frac{\partial}{\partial t} P_k^{\nu c_1}(t) &= \left[ \epsilon_k^{c_1} - \epsilon_k^{v} - i \frac{1}{T_2} + i E(t) \nabla_k \right] P_k^{\nu c_1}(t) 
 - \left( 1 - f_k^{c_1}(t) - f_k^{v}(t) \right) d^{c_1 \nu}_k E(t) \\
& \quad + E(t) \left( d^{c_2 \nu}_k P_k^{c_2 c_1}(t) - d^{c_1 c_2}_k P_k^{\nu c_2}(t) \right) \\
i \frac{\partial}{\partial t} P_k^{\nu c_2}(t) &= \left[ \epsilon_k^{c_2} - \epsilon_k^{v} - i \frac{1}{T_2} + i E(t) \nabla_k \right] P_k^{\nu c_2}(t) 
- \left( 1 - f_k^{c_2}(t) - f_k^{v}(t) \right) d^{c_2 \nu}_k E(t) \\
& \quad + E(t) \left( d^{c_1 \nu}_k P_k^{c_1 c_2}(t) - d^{c_2 c_1}_k P_k^{\nu c_1}(t) \right) \\
i \frac{\partial}{\partial t} P_k^{c_1 c_2}(t) &= \left[ \epsilon_k^{c_2} - \epsilon_k^{c_1} - i \frac{1}{T_2} + i E(t) \nabla_k \right] P_k^{c_1 c_2}(t) 
 + \left( f_k^{c_2}(t) - f_k^{c_1}(t) \right) d^{c_2 c_1}_k E(t) \\
& \quad + E(t) \left( d^{\nu c_1}_k P_k^{\nu c_2}(t) - d^{c_2 \nu }_k \left( P_k^{\nu c_1}(t) \right)^* \right)
\end{aligned}
\end{equation}

\begin{equation}\label{SBE3_f}
\begin{aligned}
\frac{\partial}{\partial t} f_k^{\nu}(t) &= -2 \text{Im} \left[ d_k^{c_1 v} E(t) \left( P_k^{\nu c_1}(t) \right)^* + d_k^{c_2 v} E(t) \left( P_k^{\nu c_2}(t) \right)^* \right] + E(t) \nabla_k f_k^{\nu}(t) \\
\frac{\partial}{\partial t} f_k^{c_1}(t) &= -2 \text{Im} \left[ d_k^{c_1 \nu} E(t) \left( P_k^{\nu c_1}(t) \right)^* + d_k^{c_1 c_2} E(t) \left( P_k^{c_2 c_1}(t) \right)^* \right]+ E(t) \nabla_k f_k^{c_1}(t) \\
\frac{\partial}{\partial t} f_k^{c_2}(t) &= -2 \text{Im} \left[ d_k^{c_2 v} E(t) \left( P_k^{\nu c_2}(t) \right)^* + d_k^{c_2 c_1} E(t) \left( P_k^{c_1 c_2}(t) \right)^* \right]+ E(t) \nabla_k f_k^{c_2}(t)
\end{aligned}
\end{equation}

Here, $P_k^{\lambda, \lambda'}(t)$ represents a dimensionless polarization depending on time $t$ and momentum $k$, induced by the time-dependent electric field $E(t)$, with $\lambda, \lambda' = \{v, c_1, c_2\}$. The indices $v$, $c_1$, and $c_2$ correspond to the valence band, first conduction band, and second conduction band, respectively. $f_k^\lambda(t)$ denotes the population of electrons or holes in the conduction or valence bands. The energy-momentum dispersion relation for the band $\epsilon_k^\lambda$, as well as the transition dipole moments (TDMs) $d_k^{\lambda \lambda'}$, are derived from density functional theory (DFT) calculations~\cite{Shicheng_PRB2020}. The exact band structure and TDMs used in our calculations are plotted in Fig.~S1. Here, we consider a "smooth-periodic" gauge such that TDMs maintain smoothness and periodicity at the boundary of the $k$-space, preserving the crystal symmetry. This ensures the absence of unphysical even-order harmonics in a single-color field from an inversion-symmetric crystal structure like MgO. We set the phenomenological decoherence time $T_2$ to half of the optical cycle duration of the 800\,nm laser. This approach cuts off long electron-hole trajectories and achieves good agreement with the experiment.

The HHG spectrum is computed as 
\begin{equation}
S_{\text{HHG}}(\omega) \sim \left| \int_{-\infty}^{\infty} [J_{\text{inter}}(t) + iJ_{\text{intra}}(t)] e^{i\omega t} dt \right|^2,
\end{equation}
where 
\begin{equation}
J_{\text{inter}}(t) = \sum_{\lambda, \lambda'} \frac{d}{dt} \int d_k^{\lambda \lambda'} \cdot P_k^{\lambda \lambda'} (t) dk + \text{c.c.}
\end{equation}
and
\begin{equation}
J_{\text{intra}}(t) = \sum_{\lambda} \int \nu_k^{\lambda}  f_k^{\lambda}(t) dk
\end{equation}
denote the interband and intraband contributions to the HHG yield, respectively. $\nu_k^{\lambda} = \nabla_k \epsilon^{\lambda}_k$ represents the group velocity of the electrons and holes in the corresponding bands.

\subsection*{Details of the experimental setup}

The two-color delay is controlled by a closed-loop linear piezo stage (Physik Instrumente P-611.1). For combining the two colors, we use a dichroic mirror (Thorlabs DMLP1180T). Starting with the focusing optics, the HHG setup is situated in an ultra-high vacuum chamber with a base pressure of $\sim 1 \times 10^{-8}$\,mbar. The laser beam is focused using an off-axis parabolic mirror with a focal length of 5\,cm. The spot size of the 800\,nm beam at the focus is $\sim$6\,\(\mu\)m FWHM, as measured by a camera. The 2000\,nm beam at the focus has an effective radius of $\sim$18.2 \(\mu\)m, as determined by a knife-edge measurement. The pulse durations were measured using frequency-resolved optical gating. The $<100>$-cut MgO crystal used in our experiment has a thickness of $50\,\mu$m (Farview Optics). The home-built spectrometer uses a $100\,\mu m$ wide entrance slit and a flat-field diffraction grating suitable for wavelengths of $50-200$\,nm (L0120-50-200, Shimadzu).

We also determine the divergence angle of the generated harmonics by recording the \(1/e^2\) width at the MCP position and the distance between the MCP detector and the MgO crystal. H11 has a divergence angle of approximately 17.9\,mrad. With a pump intensity of $15\,\mathrm{TW\,cm}^{-2}$, we find that the lower bound of the photon flux of the measured HHG signal is on the order of $10^7$ photons per second in the spectral range from 10\,eV to 20\,eV. For this estimate, we take into account the grating and the low detection efficiency of our uncoated MCP (Texel MCP-50-D-S-P43).


\clearpage 

%

\bibliographystyle{sciencemag}

%
%
%
%
%
%


\section*{Acknowledgments}
The authors acknowledge Barry D.~Bruner and Nirit Dudovich for insightful discussions and Ulf Leonhardt and Lorenzo M.~Procopio for providing equipment for pulse characterization. 

\paragraph*{Funding:}
This project has received funding from the European Union's Horizon 2020 research and innovation program under grant agreement No 853393-ERC-ATTIDA. We also acknowledge the Helen Diller Quantum Center at the Technion for partial financial support.

\paragraph*{Author contributions:}
M.~K.~and Z.~C.~conceived the initial idea. M.~K.~supervised the project. M.~L., Z.~C., B.~R.~and M.~K.~designed and built the experimental setup. M.~L., Y.~K.~and Z.~C.~performed the measurements. Z.~C.~and A.~G.~supported the operation of the laser system. Z.~C.~carried out the theory calculations. Z.~C.~and M.~K.~wrote the initial manuscript. All the authors contributed to the preparation of the final manuscript.

\paragraph*{Competing interests:}
The authors declare that they have no competing interests.

\paragraph*{Data and materials availability:}
All data needed to evaluate the conclusions in the paper are present in the paper and/or the Supplementary Materials.




\newpage


\renewcommand{\thefigure}{S\arabic{figure}}
\renewcommand{\thetable}{S\arabic{table}}
\renewcommand{\theequation}{S\arabic{equation}}
\renewcommand{\thepage}{S\arabic{page}}
\setcounter{figure}{0}
\setcounter{table}{0}
\setcounter{equation}{0}
\setcounter{page}{1} 


\begin{center}
\section*{Supplementary Materials for\\ \scititle}

Zhaopin Chen \textit{et al}.\\
Corresponding authors. Email: zhaopin.chen@campus.technion.ac.il, krueger@technion.ac.il
\end{center}



\newpage

\section*{Supplementary text}\label{sec0}

\subsection*{1. Band structure and transition dipole moment of MgO for numerical calculations}

The three-band structure and transition dipole moment (TDM) of MgO along the $\Gamma-X$ orientation in our numerical simulation is obtained from a density functional theory calculation~\cite{Shicheng_PRB2020}. We also introduce a small modification of the TDM amplitude to better fit the experimental results. We show the band structure and TDM in Fig.~S1.

\subsection*{2. Semiconductor Bloch equation simulation results for CEP equal to zero }

Here, we present additional simulation results of the two-color control of high harmonic generation from MgO, based on the semiconductor Bloch equations (1) and (2), using the same parameters as those in Fig.~3 of the main text. Figure S2A displays the harmonic spectrum as a function of the two-color delay, ranging from $-$120\,fs to 120\,fs. Similar to the experimental results in Fig.~2A in the main text, the spectrum shows a quasi-continuous profile in the central overlap region of the two-color pulses. Spectral interference shifts as a function of delay are also clearly observed in the simulation. The white dashed line indicates a slope of $6.5\,\mathrm{fs\,eV}^{-1}$, consistent with the experimental observations. In the Fourier time domain, Fig.~S2b reveals a suppression of the secondary peak at a Fourier time of $t \approx 1.33\,\mathrm{fs}$ and the appearance of weak intensity at $t \approx 4\,\mathrm{fs}$, closely resembling the experimental data in Fig.~2D of the main text.

In order to see the harmonic distribution in both time and frequency domain for a specific delay, we perform a Gabor transformation over the interband and intraband currents obtained by Eqs. (3) and (4) in the main text. The Gabor transformation is calculated by

\begin{equation}
G(\omega, t') = \left| \int \left[ J_\text{inter}(t) + J_\text{intra}(t) \right] \exp\left( -\frac{(t - t')^2}{\sigma_t^2} \right) e^{i \omega t} \, dt \right|^2. \label{Gabor_eq}
\end{equation}

Here, we use a Gaussian window function with a width of \(\sigma_{t} = 800 \, \text{as}\). Figure \ref{fig_S7}A clearly shows the generation of an isolated pulse with a spectrum ranging from 13\,eV to 20\,eV at a two-color delay time of \(\tau = 0 \, \text{fs}\). At \(\tau = 2.5 \, \text{fs}\), two attosecond pulses with a temporal separation of 4\,fs are produced. It is evident that the attochirp is minimal in both cases, primarily due to the contribution of the short trajectory.

\subsection*{3. Semiconductor Bloch equation simulation results for CEP equal to $\pi /2$}

In order to assess the influence of CEP of both pulses on the experimental results, we consider a case with $\phi_\mathrm{CEP,1}=\phi_\mathrm{CEP,2}=\pi/2$, which is the opposite with respect to Fig.~3 in the main text with zero CEP phase. Even with $\pi/2$ phase for 800\,nm where the positive and negative peak intensity are equal in the center of the pulse, a high intensity ratio between the dominant peak and secondary peak is still obtained, $\sim 79\%$ (see Fig.~S4A). The corresponding attosecond pulse distribution from the generated high harmonics is highly similar to the one with $\phi_{CEP,1}=\phi_{CEP,2}=0$ in Fig.~3C in the main text. In this case, an isolated attosecond pulse with high contrast can still be realized under specific delay, e.g., $\tau=0.9\, \text{fs}$ (see Fig.~S4B). The spectral intensity as a function of two-color delay in Fig.~ S4 is very similar to the case of zero CEP. Our results show that the CEP does not affect the main physics and behaviours in our two-color HHG experiments, except for a small influence on the IAP duration ($\sim 100$\,as). 

\subsection*{4. Quantum efficiency of the micro-channel plate detector}

Figure~\ref{MCP} shows the estimated quantum efficiency of the micro-channel plate (MCP) detector used to measure high harmonics in the experiment. To simulate the experimental results, we also multiplied the SBE simulation spectrum by the MCP efficiency.

\subsection*{5. Comparison of harmonic spectrum from intra- and interband currents }

Figure \ref{J_inter_intra} compares the harmonic spectral intensity contributions from intraband and interband currents driven by an 800\,nm pulse with a peak intensity of $15\,\mathrm{TW}\,\mathrm{cm}^{-2}$. Strikingly, the harmonic intensity generated by the interband current is several orders of magnitude higher than that from the intraband current.

\newpage


\begin{figure}[htb!]
\centering
\includegraphics[width=0.9\textwidth]{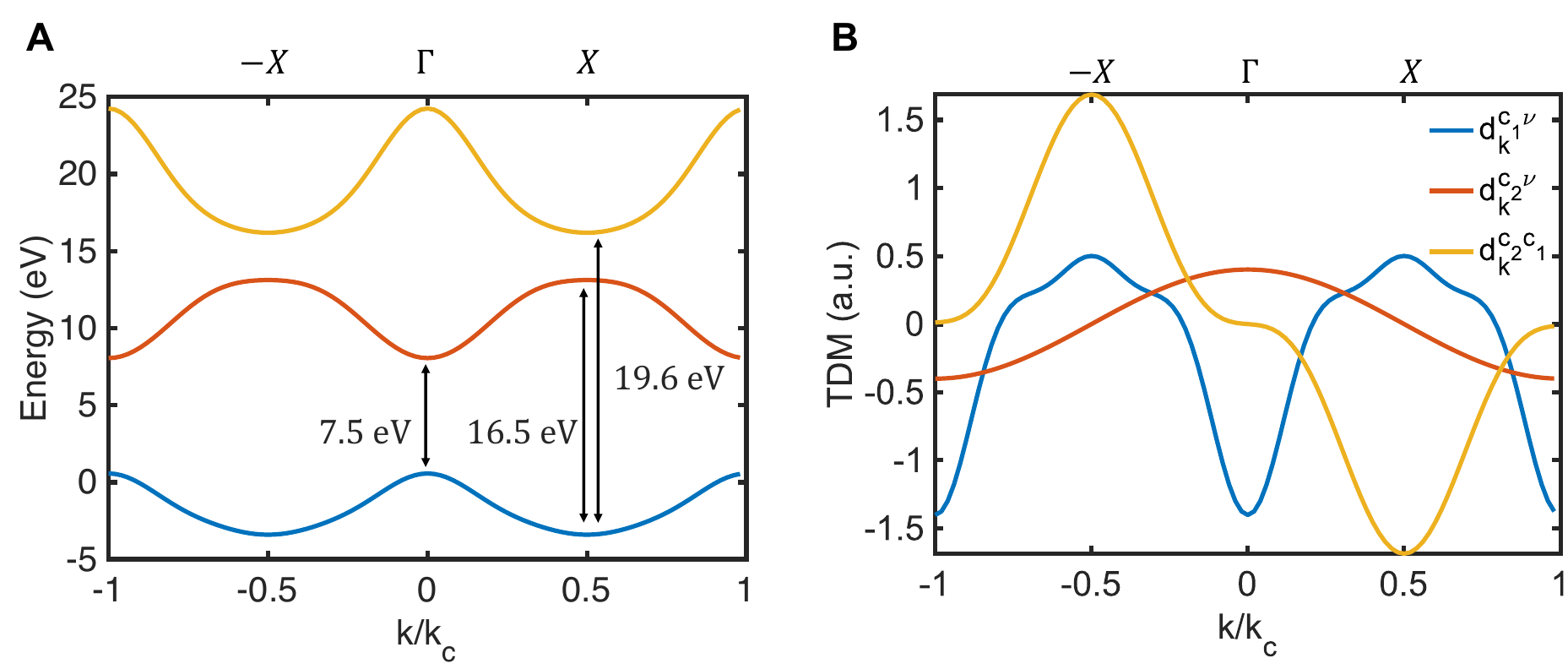} 
\caption{ {\bf Band structure and transition dipole moment of MgO}. ({\bf A}) Band structure of the MgO crystal in the $\Gamma-X$ orientation. The bottom curve is the valence band and the upper two curves represent the first and second conduction bands, respectively. ({\bf B}) Corresponding transition dipole moments (TDMs) in atomic units (a.u.). }\label{fig_S1}
\end{figure}

\newpage

\begin{figure}[htb!]
\centering
\includegraphics[width=1\textwidth]{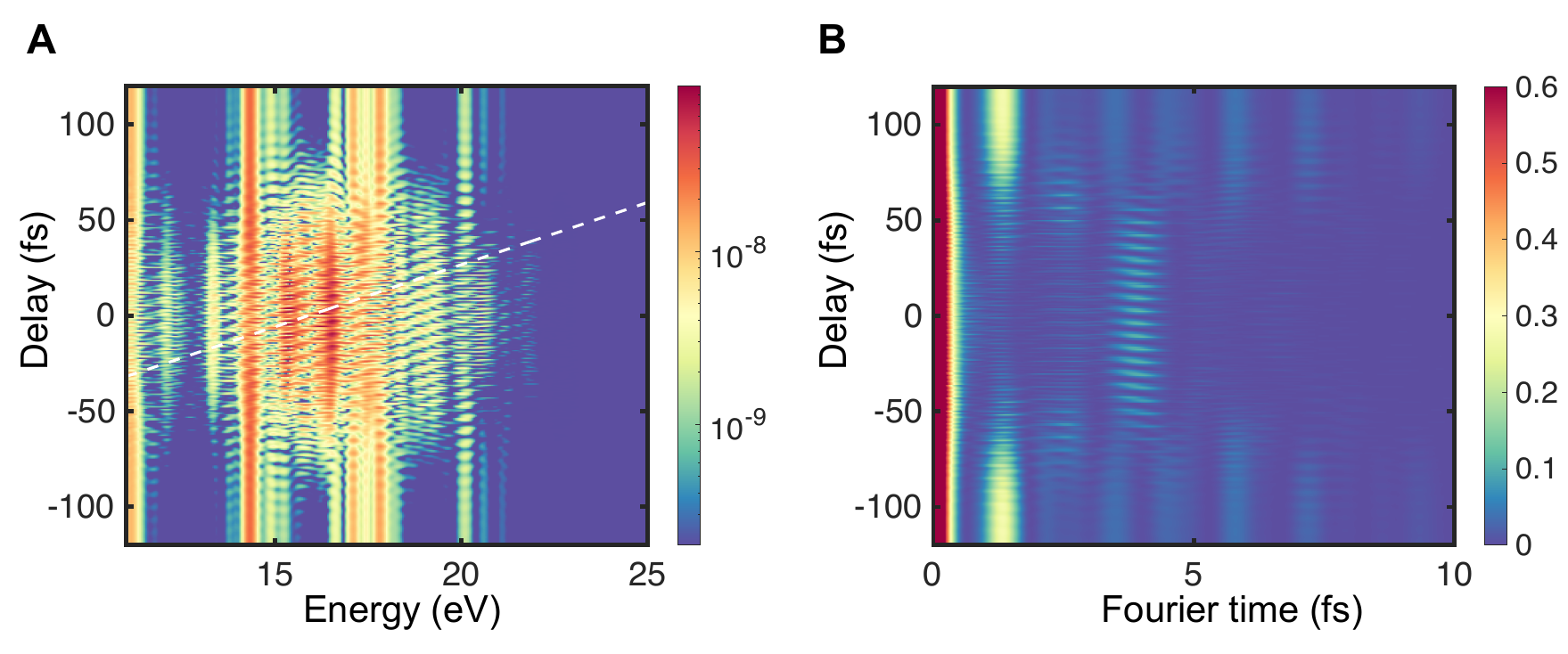}
\caption{{\bf Harmonics intensity in spectral and Fourier time domain}. ({\bf A})Harmonic spectrum as a function of two-color delay from $-120\, \mathrm{fs}$ to $120\, \mathrm{fs}$ calculated with the SBEs. The white dashed line has a slope of $6.5\,\mathrm{fs}\,\mathrm{eV}^{-1}$. ({\bf B}) Harmonic intensity in the Fourier time domain as a function of delay, corresponding to ({\bf A}). }\label{fig_S2}
\end{figure}

\newpage

\begin{figure}[htb!]
\centering
\includegraphics[width=1\textwidth]{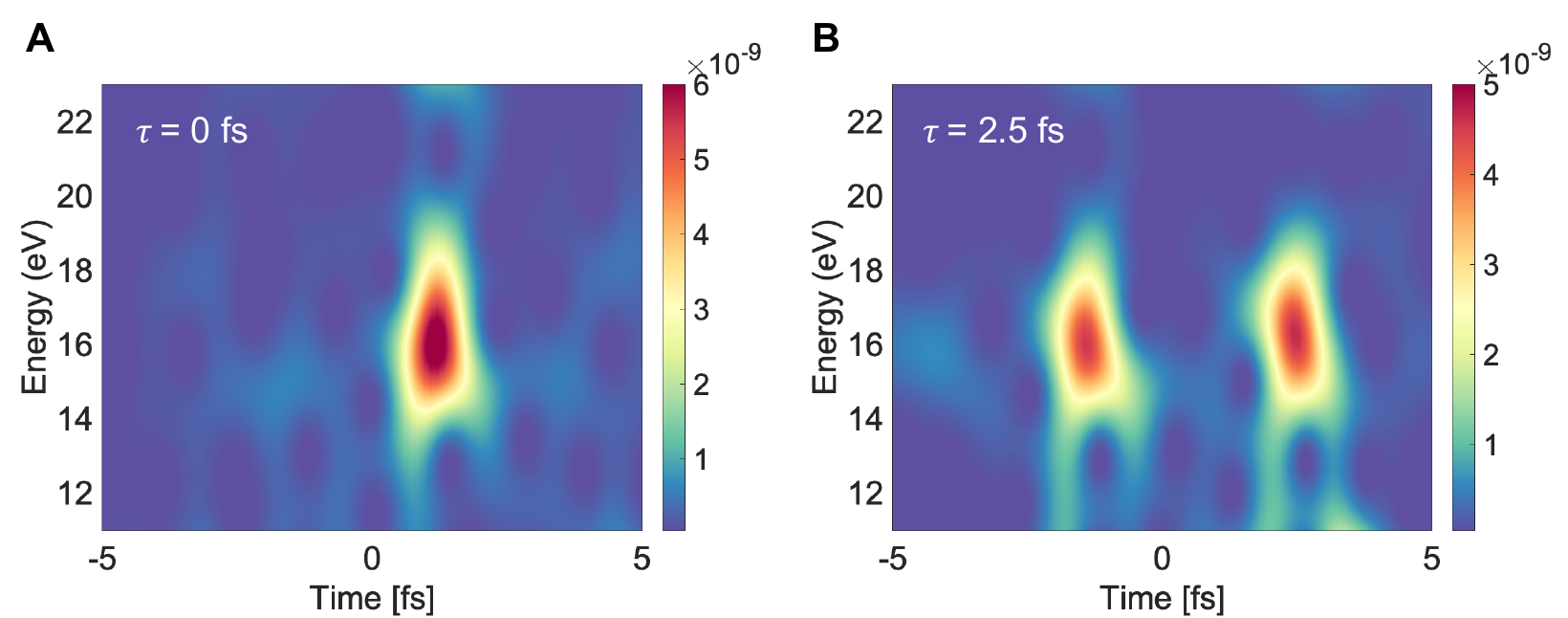}
\caption{{\bf Time-frequency analysis}. Gabor transformation of the simulated harmonics in Fig.~3 of the main text, based on the definition in Eq.~\ref{Gabor_eq}. ({\bf A}) and ({\bf B}) correspond to two-color delays $\tau=0\,\mathrm{fs}$ and $\tau=2.5 \,\mathrm{fs}$, respectively.}\label{fig_S7} 
\end{figure}

\newpage

\begin{figure}[htb!]
\centering
\includegraphics[width=1\textwidth]{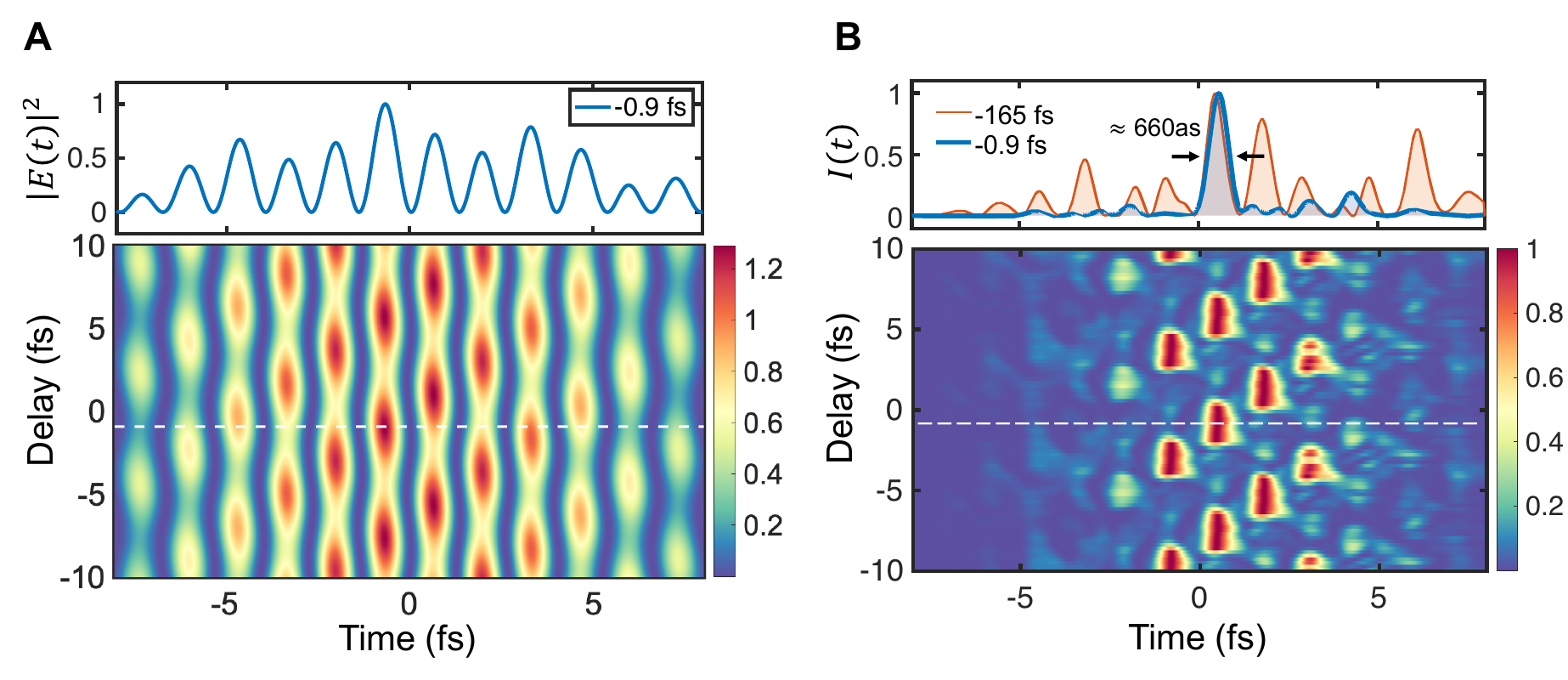}
\caption{ {\bf Time-domain simulation results for CEP equal to $\pi/2$}. ({\bf A}) Field strength square $|E(t)|^{2}$ distribution of the synthesized pulse as a function of the pulse delay. The upper plot shows the synthetic field distribution in the white dash line with two-color delay equal to $-0.9\, \mathrm{fs}$. The intensity ratio between the dominant peak and secondary peak is $\sim 79\%$. ({\bf B}) Attosecond pulses intensity as a function of two-color delay in real time. The upper plot compares the formation of attosecond pulse trains and isolated attosecond pulses at time delay $-165\, \mathrm{ fs}$ (orange) and $-0.9\, \mathrm{fs}$ (blue), respectively. The dash line in the lower plot indicates the corresponding time delay. Similar to Fig.~S2, the peak intensity for the 800\,nm pulse is $\mathrm{15\,TW\,cm^{-2}}$, but with the CEP $\phi_\mathrm{CEP}=\pi/2$.}\label{fig_S3}
\end{figure}

\newpage

\begin{figure}[htb!]
\centering
\includegraphics[width=1\textwidth]{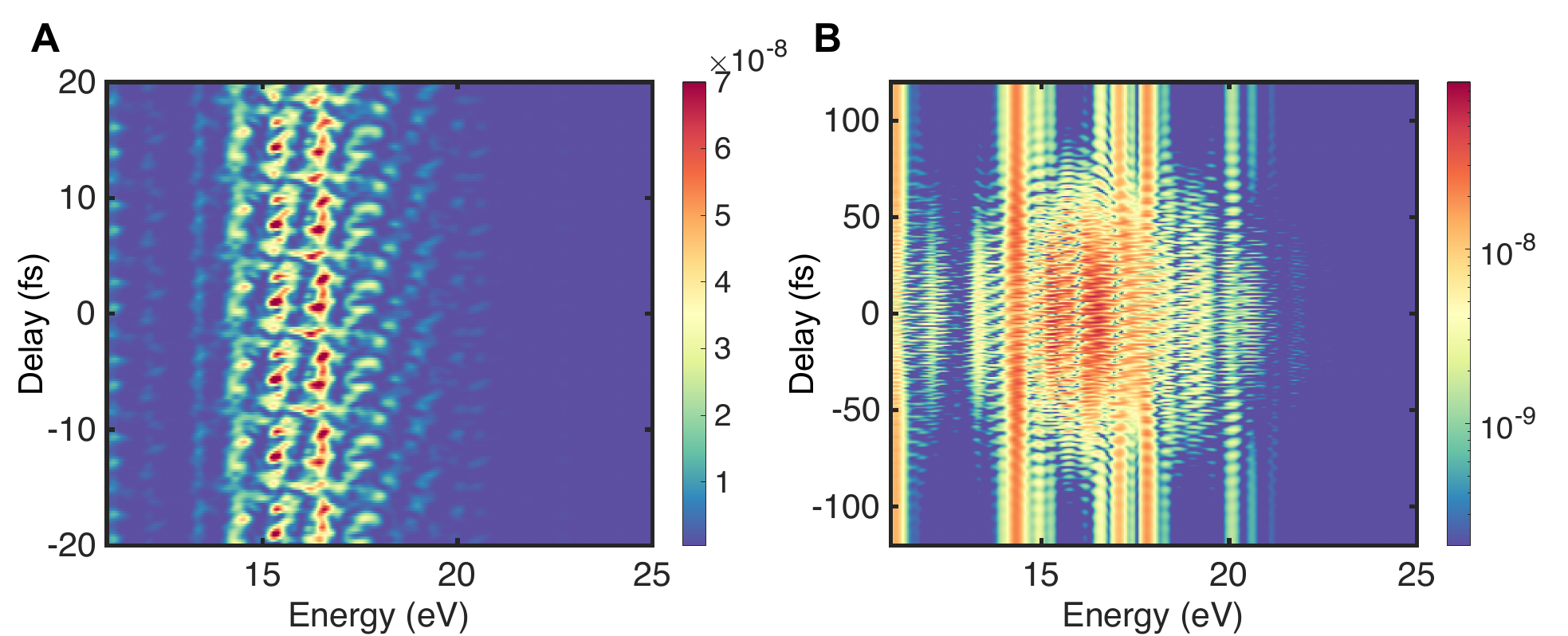}
\caption{{\bf Frequency-domain simulation results for CEP equal to $\pi/2$}. Harmonic intensity as a function of two-color delay for SBE simulation for ({\bf A}) range from $-20\,\mathrm{fs}$ to $20 \,\mathrm{fs}$, and ({\bf B}) from $-120\,\mathrm{fs}$ to $120\,\mathrm{fs}$. Similar to Fig.~S3, the peak intensity for the 800\,nm pulse is $\mathrm{15\,TW\,cm^{-2}}$, but with CEP phase $\phi_\mathrm{CEP}=\pi/2$. }\label{fig_S4} 
\end{figure}

\newpage

\begin{figure}[htb!]
\centering
\includegraphics[width=0.5\textwidth]{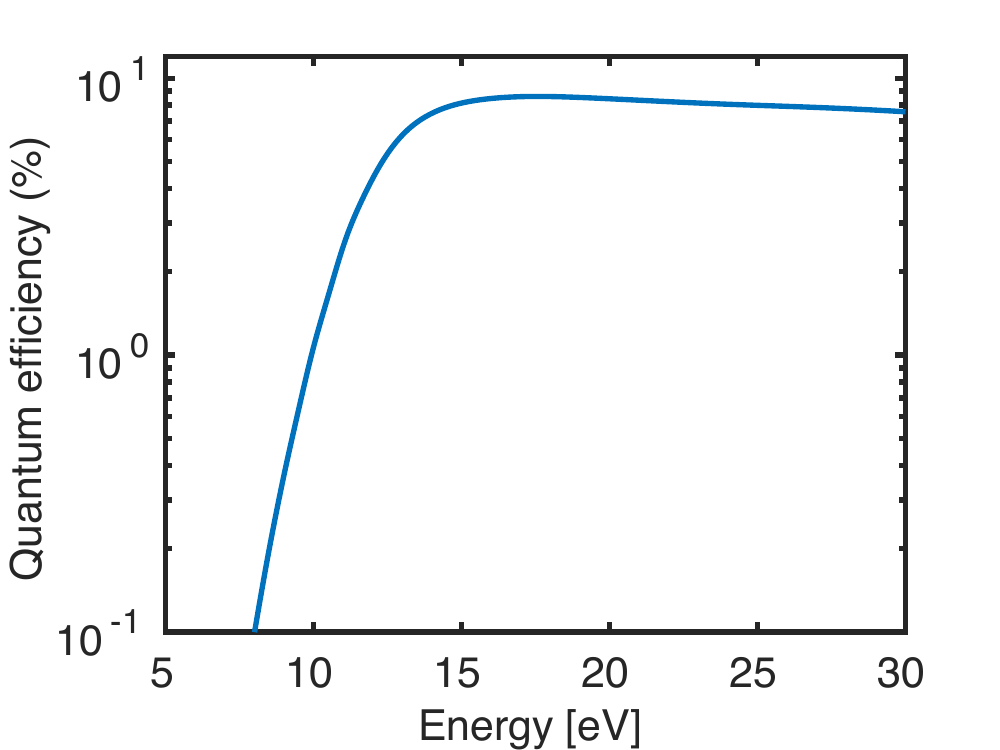}
\caption{{\bf MCP quantum efficiency}. Estimated quantum efficiency of the microchannel plate detector as a function of photon energy. For the SBE simulation results in this paper, we account for this efficiency to model the experimental results. }\label{MCP} 
\end{figure}

\newpage

\begin{figure}[htb!]
\centering
\includegraphics[width=0.5\textwidth]{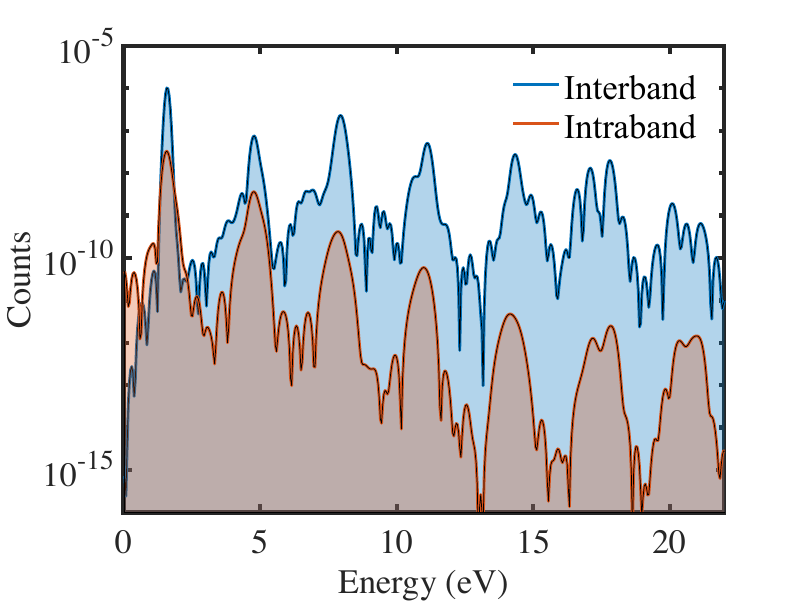}
\caption{{\bf Harmonics spectrum from interband and intraband currents}. Comparison between interband and intraband contribution to the harmonic spectrum for a single-color 800\,nm pump with $15 \mathrm{TW}\,\mathrm{cm}^{-2}$. }\label{J_inter_intra} 
\end{figure}

\newpage






\end{document}